\documentclass[conference]{IEEEtran}
\IEEEoverridecommandlockouts
\usepackage[utf8]{inputenc}
\usepackage[T1]{fontenc}
\usepackage[font=small]{caption}
\usepackage{cite}
\usepackage{graphicx}
\usepackage[cmex10]{amsmath}
\usepackage{algorithmic}
\usepackage{url}
\usepackage[utf8]{inputenc}
\usepackage{amssymb,amsfonts}
\usepackage{textcomp}
\usepackage{subfigure}
\usepackage{xcolor}
\usepackage[export]{adjustbox}
\usepackage{comment}
\usepackage{paralist}
\usepackage{enumitem}
\usepackage[bookmarks=false, hidelinks]{hyperref}
\usepackage[T1]{fontenc}
\usepackage{tabularx}
\usepackage{multirow}
\usepackage{makecell}
\usepackage{balance}
\usepackage{hhline}
\usepackage{comment}

\begin{document}

\title{Resilience-driven Planning of Electric Power Systems Against Extreme Weather Events}

\author{\IEEEauthorblockN{Abodh Poudyal$^\dagger$, Shishir Lamichhane, and Anamika Dubey}
\IEEEauthorblockA{Washington State University, Pullman, WA, USA\\
Email: $^\dagger$abodh.poudyal@wsu.edu}\vspace{-1cm}
\and  
\IEEEauthorblockN{Josue Campos do Prado}
\IEEEauthorblockA{Washington State University, Vancouver, WA, USA}} 

\maketitle
\thispagestyle{plain}
\pagestyle{plain}

\begin{abstract}
With the increasing frequency of natural disasters, operators must prioritize improvements in the existing electric power grid infrastructure to enhance the resilience of the grid. Resilience to extreme weather events necessitates lowering the impacts of high-impact, low-probability (HILP) events, which is only possible when such events are considered during the planning stage. This paper proposes a two-stage stochastic planning model where the generation dispatch, line hardening, line capacity expansion, and distributed generation sizing and siting decisions are proactively decided to minimize the overall load shed and its risk for extreme weather scenarios, where the risk is modeled using conditional value-at-risk. To alleviate computational complexity without sacrificing solution quality, a representative scenario sampling method is used. Finally, the overall framework is tested on a standard IEEE reliability test system to evaluate the effectiveness of the proposed approach. Several planning portfolios are presented that can help system planners identify trade-offs between system resilience, planning budget, and risk aversion. 
\end{abstract}

\begin{IEEEkeywords}
power system resilience, risk-based planning, stochastic optimization, high-impact low-probability events
\vspace{-0.2cm}
\end{IEEEkeywords}

\section{Introduction}\label{sec:intro}

In recent years, the frequency of severe weather events has risen significantly as a result of climate change. The annual average number of disasters exceeding \$1 billion from 1980 to 2022 is 8.1, whereas the most recent five years, from 2018 to 2022, has an average of 18~\cite{NOAA2023}. Approximately 83\% of all major outages reported in the U.S. between 2000 and 2021 were weather-related~\cite{climatecentral2022}. 
The aftermath of such events incurs a significant amount of socio-economic losses to the end users and an economic burden to the grid operators. 
Hence, there is a strong urgency for grid operators to have resilience considerations during planning due to the growing frequency of extreme weather events.

There are several works focusing on resilience enhancement transmission planning. In~\cite{ranjbarresiliency}, a two-stage stochastic planning model is proposed for transmission line upgrade and DERs investment. The study presented in~\cite{datadriven} introduces a data-driven planning model for transmission systems. This model aims to identify the most effective portfolio for enhancing resilience against extreme weather events in the power system. However, the work only focuses on worst-case scenarios. A resilient network investment model that co-optimizes substation hardening and transmission expansion to protect the system against extreme events is proposed in~\cite{substation_line_coOptimized}. Resilience-based planning of distributed series reactors is proposed in~\cite{reactorUsed}. The planning method particularly focuses on extreme weather events. All of these works consider the DC optimal power flow (DCOPF) model, which is a decent approximation model for ACOPF. However, DCOPF does not necessarily have resilience considerations when the generator dispatch decisions are made. The work in~\cite{bynum2021proactive} addresses this issue by formulating a resilience-driven proactive dispatch model. However, the approach only considers a targeted number of outage scenarios, which is inappropriate for planning purposes.
For instance, ten different line outages might incur no loss in a meshed network. However, the outage of even two lines connecting the generator to the load can instigate a blackout in the system. Additionally, none of the above-mentioned works consider risk minimization and only focus on expected outcomes, which often fail to characterize and incorporate system resilience within the optimization.  

This work proposes a resilience-driven two-stage stochastic planning model where generation proactive dispatch, line hardening, line capacity upgrading, DG siting, and sizing decisions are decided to minimize the expected load loss and generation curtailment while minimizing the risk of load loss. The risk is modeled using conditional value-at-risk (CVaR), which has been widely used in other risk-based planning models~\cite{poudyal2022risk}. 
A regional wind profile represents the extreme wind storm event, and the impact on the grid is observed using component-level fragility curves. 
The damage scenarios are generated using Monte-Carlo simulation, and to alleviate the computational complexity, a strategic representative scenario selection method is utilized as described in~\cite{poudyal2022risk}. 
Finally, a two-stage risk-averse stochastic problem is formulated where the first stage involves planning and proactive generator dispatch decisions, whereas the second stage facilitates optimal operation of the power grid based on the planning decision for every scenario. 
The overall contributions of this paper are:

\begin{enumerate}
    \item \textit{A risk-based resilience planning model for electric power systems:} We propose a two-stage risk-based mixed integer linear programming (MILP) model for resilience planning in the power grid. The planning decisions include proactive generator dispatch decisions, line hardening, line capacity upgrade, and DG siting and sizing. The effectiveness of the proposed resilience-driven planning model is tested on the IEEE Reliability Test System~\cite{barrows2019ieee} 
    \item \textit{Trade-off analysis on multiple planning portfolios:} A comprehensive trade-off analysis is presented on several resource portfolios compared to the planning budget and risk aversion. The grid operators can effectively utilize the trade-off in making planning decisions based on the planning objective and availability of the budget and resources.
\end{enumerate}

The remainder of the manuscript is organized as follows. Section~\ref{sec:modeling} provides the mathematical details of the proposed resilience-driven two-stage stochastic planning model. Section~\ref{sec:scenario} describes the overall scenario generation and selection followed by results and analysis in Section~\ref{sec:result} with conclusion in Section~\ref{conclusion}.

\section{Resilience-driven Planning Modeling}\label{sec:modeling}
\subsection{Two-stage Stochastic Planning Model}
The DCOPF model obtains the cost-economical generation dispatch decisions. However, DCOPF-based dispatch decisions do not consider system resilience and future extreme events within its model. Furthermore, once the dispatch decisions are identified, generators are constrained by their ramp rates, which makes it challenging to re-dispatch in case of extreme weather events.
This proposed two-stage stochastic planning model provides generation dispatch, and investment planning decisions to minimize the total load shed and risk due to extreme weather scenarios. The overall planning model is shown below:  

\begin{equation}
\small
    \begin{split}
    \min (1-\lambda)\mathbb{E}\left(\mathcal{F}(\textbf{x}, \mathcal{E}) + \frac{\gamma \hat{P}^\mathcal{E}_{\mathcal{G}}}{1 - \lambda}\right) + \lambda CVaR_\alpha(\mathcal{F}(\textbf{x}, \mathcal{E}))
    \end{split}
    \label{eq:stochastic_objective}
\end{equation}
where,
\begin{equation*}
\small
    \begin{gathered}
        \textbf{x} \in \{x^h, x^u, x^{DG}, \zeta^{DG}, P_{\mathcal{G}}\}\\
        \mathbb{E}(\mathcal{F}(\textbf{x}, \mathcal{E}) + \gamma \hat{P}^\mathcal{E}_{\mathcal{G}}) := \sum_{\xi \in \mathcal{E}} p^\xi \left(\sum_{i \in \mathcal{B}}\hat{P}_i^{D,\xi} + \gamma \sum_{i_G \in \mathcal{G}} \hat{P}_{i_G}^\xi)\right) \\
        CVaR_\alpha(\mathcal{F}(\textbf{x}, \mathcal{E})) := \left(\eta + \frac{1}{1-\alpha}\sum_{\xi \in \mathcal{E}}p^\xi\nu^\xi\right)
    \end{gathered}
\end{equation*}


\noindent
subject to,

\begin{equation}
\small
\begin{gathered}
    \sum_{i_G \in \mathcal{G_B}}P_{i_G} + \sum_{e:(j,i)\in \mathcal{L}} P_e - \sum_{e:(i,j)\in \mathcal{L}} P_e = P_i^D + G_i\\ \forall i,j \in \mathcal{B}, \mathcal{G_B} \in \mathcal{G}    
\end{gathered}
    \label{eq:power_balance}
\end{equation}

\begin{equation}
    P_e = \frac{B_e}{N_e^T}\left(\theta_i - \theta_j - \Phi_e^T \right)~\forall e:(i,j)\in \mathcal{L}~i,j\in \mathcal{B} 
    \label{eq:power_flow}
\end{equation}

\begin{equation}
    P_{i_G}^{min} \leq  P_{i_G} \leq  P_{i_G}^{min}~\forall i_G \in \mathcal{G}
    \label{eq:generation_limit}
\end{equation}

\begin{equation}
    -P_{e}^{max} \leq P_{e} \leq  P_{e}^{max}~\forall e:(i,j)\in \mathcal{L} 
    \label{eq:flow_limit}
\end{equation}

\begin{equation}
    -\theta_{e}^{max} \leq \theta_i - \theta_j \leq \theta_{e}^{max}~\forall e:(i,j)\in \mathcal{L},~i,j\in \mathcal{B}  
    \label{eq:angle_limit}
\end{equation}

\begin{equation}
\begin{split}
     \sum_{e\in \mathcal{L}} c_e^{h}x_e^{h} + \sum_{e\in \mathcal{L}} c_e^{u}x_e^{u} + \sum_{i\in \mathcal{B}} c_{i}^{DG}x_{i}^{DG}     
     \leq \mathcal{C}^{max}_{T}
     \\~\forall e:(i,j)\in \mathcal{L}
\end{split}
    \label{eq:overall_budget}
\end{equation}

\begin{equation}
    \sum_{e\in \mathcal{L}} x_e^{h} \leq N_h, \sum_{e\in \mathcal{L}} x_e^{u} \leq N_u, \sum_{i\in \mathcal{B}} \zeta_i^{DG} \leq N_{DG}~\forall e:(i,j)\in \mathcal{L}
    \label{eq:investment_number_constraint}
\end{equation}

\begin{equation}
   0 \leq  x_i^{DG} \leq  \zeta_i^{DG} P_{i}^{DG, max}~\forall i \in \mathcal{B}
    \label{eq:DG_limit}
\end{equation}

\begin{equation}
    - P_{i_G}^R \leq  P_{i_G} - P_{i_G}^\xi \leq P_{i_G}^R~\forall i_G \in \mathcal{G},~\forall \xi \in \mathcal{E}
    \label{eq:stochastic_ramp_limit}
\end{equation}

\begin{equation}\label{eq:stochastic_power_balance}
    \small
    \begin{gathered}
    \sum_{i_G \in \mathcal{G_B}}(P_{i_G}^\xi - \hat{P}_{i_G}^\xi) + \sum_{i \in \mathcal{B}} x_i^{DG} + \sum_{e:(j,i)\in \mathcal{L}} P_e^\xi - \sum_{e:(i,j)\in \mathcal{L}} P_e^\xi = \\ P_i^D + G_i - \hat{P}_i^{D,\xi} ~\forall i,j \in \mathcal{B},~\forall \xi \in \mathcal{E}, \mathcal{G_B} \in \mathcal{G}
    \end{gathered}
\end{equation}

\begin{equation}\label{eq:stochastic_connectivity}
    \delta_e^\xi = \pi_e^\xi + x_e^{h} - \pi_e^\xi \times x_e^{h}~\forall e:(i,j)\in \mathcal{L},~\forall \xi \in \mathcal{E}
\end{equation}

\begin{equation}
\small
\begin{gathered}
     -(1 - \delta_e^\xi)\textbf{M} \leq \frac{P_e^\xi \times N_e^T}{B} - \left(\theta_i^\xi - \theta_j^\xi - \phi_e^T \right) \leq (1 - \delta_e^\xi)\textbf{M}\\~\forall e:(i,j)\in \mathcal{L},~i,j\in \mathcal{B},~\forall \xi \in \mathcal{E} 
    \label{eq:stochastic_power_flow}
\end{gathered}
\end{equation}

\begin{equation}
\small
\begin{gathered}
    -\delta_e^\xi(1 - x_e^u) P_{e}^{max} - (1 + \phi_e^u) x_e^u \delta_e^\xi P_{e}^{max} \leq P_{e}^\xi \leq  \delta_e^\xi(1 - x_e^u) P_{e}^{max} \\ + (1 + \phi_e^u) x_e^u \delta_e^\xi P_{e}^{max}~\forall e:(i,j)\in \mathcal{L},~\forall \xi \in \mathcal{E} 
\end{gathered}
    \label{eq:stochastic_flow_limit}
\end{equation}

\begin{equation}
\small
\begin{gathered}
    -\theta_{e}^{max} - \textbf{M}(1 - \delta_e^\xi) \leq \theta_i^\xi - \theta_j^\xi \leq \theta_{e}^{max} + \textbf{M}(1 - \delta_e^\xi)\\~\forall e:(i,j)\in \mathcal{L},~i,j\in \mathcal{B},~\forall \xi \in \mathcal{E}  
\end{gathered}
    \label{eq:stochastic_angle_limit}
\end{equation}

\begin{equation}
    P_{i_G}^{min} \leq  P_{i_G}^\xi \leq  P_{i_G}^{max}~\forall i_G \in \mathcal{G},~\forall \xi \in \mathcal{E}
    \label{eq:stochastic_generation_limit}
\end{equation}

\begin{equation}
     0 \leq \hat{P}_i^{D,\xi} \leq P_i^D
    \label{eq:load_shed_limit}~\forall i \in \mathcal{B},~\forall \xi \in \mathcal{E}
\end{equation}

\begin{equation}
     0 \leq \hat{P}_{i_G}^{\xi} \leq P_{i_G}^\xi
    \label{eq:gen_curtailment_limit}~\forall i_G \in \mathcal{G},~\forall \xi \in \mathcal{E}
\end{equation}

\begin{equation}
    \nu^{\xi} \geq \mathcal{F}(\textbf{x}, \xi) - \eta, \hspace{0.3cm} \nu \in \mathbb{R}^n_+, ~\forall \xi \in \mathcal{E} 
    \label{eq:CVaR_constraint}
\end{equation}

\subsubsection{Objective Function}
The objective function in (\ref{eq:stochastic_objective}) minimizes the weighted sum of expected load shed ($\hat{P}_i^{D,\xi}$) and CVaR and an additional generation curtailment ($\hat{P}_{i_G}^\xi$) for a set of extreme weather event scenarios, $\xi \in \mathcal{E}$.
The generation curtailment term, weighed by $\gamma$, is introduced to maintain a feasible power flow solution in cases when there are islands with more generation than load. The value of $\gamma$ is set very low to ensure higher priority to minimize the weighted sum of expected load shed and CVaR. CVaR has been widely adopted as a risk metric for resilience planning and quantification in power systems and can be easily integrated with stochastic models~\cite{poudyal2022risk} and is described in detail in~\cite{rockafellar2000optimization}. The overall planning decisions are line hardening decision ($x^h$), line capacity expansion decision ($x^u$), DG sizing ($x^{DG}$) and siting ($\zeta^{DG}$) decisions, and proactive generator dispatch decision ($P_{\mathcal{G}}$). In (\ref{eq:stochastic_objective}), $\mathcal{B}$ represents the set of buses, $\mathcal{L}$ represents the set of lines, $\mathcal{G}$ represents the set of generators, $\mathcal{G_B}$ represents the set of generators in bus $\mathcal{B}$, $\eta$ is the value-at-risk (VaR), $\nu^\xi$ is the excess CVaR term for each scenario $\xi$, $p^\xi$ is the probability of each scenario, and $\alpha$ is the confidence level for assessing CVaR. 
The weighing term $\lambda \in [0,1]$ represents the risk aversion of the system planner, which increases monotonically with $\lambda$. 

\subsubsection{First-stage Constraints}
Eqs. (\ref{eq:power_balance}) - (\ref{eq:angle_limit}) represent first-stage DCOPF constraints for identifying generator dispatch decisions to minimize (\ref{eq:stochastic_objective}) for all scenarios. Constraint (\ref{eq:power_balance}) ensures the power balance in each bus such that the generation and incoming power on the bus are equal to the outgoing power, demand ($P_i^D$), and shunt conductance ($G_i$) on the same bus $i \in \mathcal{B}$. The line flow is maintained by (\ref{eq:power_flow}), which is the product of line susceptance $B_e = {x_e}^{-1}$  and angle ($\theta)$ difference between the two buses connected by the line. For transformers, the line flow is also guided by their turn ratio ($N^T_e$) or phase shift angle ($\Phi^T_e$) as shown in (\ref{eq:flow_limit}). Constraints (\ref{eq:generation_limit}) - (\ref{eq:angle_limit}) bound the active power flow from the generators ($\mathcal{P_{i_G}}$), active power flow on the lines($P_e$) based on their thermal limits ($P^{max}_e$), and angle difference between two buses in a line respectively. For economic and physical operating reasons, some generators should have a minimum generation level below which they cannot operate. Hence, $P_{i_G}^{min}$ ensures that for such generators, a minimum generation level is always maintained if operated. Eqs. (\ref{eq:overall_budget}) - (\ref{eq:DG_limit}) represent constraints for first-stage investment decisions. Constraint (\ref{eq:overall_budget}) ensures that the investment on hardening, capacity upgrade, and DGs do not exceed the maximum planning budget of $\mathcal{C}^{max}_{T}$, (\ref{eq:investment_number_constraint}) restricts the maximum number of investment for each investment decision, and (\ref{eq:DG_limit}) constraints the size of DG unit that can be placed in each bus. In this work, the amount of connected bus load and utilization factor of DG units guide the maximum capacity of the DG unit at that bus. 

\subsubsection{Second-stage Constraints}
The second-stage problem is solved for each scenario, and hence, the second-stage variables are parameterized by $\xi$ throughout this paper. Constraint (\ref{eq:stochastic_ramp_limit}) restricts the generation re-dispatch within its maximum and minimum ramping ability based on the first stage proactive dispatch. Generators are limited in terms of active power modulation by their ramp rate, which is usually based on their maximum and available capacity once they are dispatched. Therefore, it is desired to carefully decide the optimal generation dispatch to enhance the system's resilience. The power balance at each bus for each scenario is maintained by (\ref{eq:stochastic_power_balance}), where generation curtailment and load shedding variables are introduced. Here, $x^{DG}_i$ supplies additional generation to minimize the load shed. Constraint (\ref{eq:stochastic_connectivity}) decides the line connectivity based on the outage and hardening decision. Here, $\pi_e^\xi$ is an exogenous parameter and is known when a scenario is realized such that when $x^h_e = 0$, $\delta_e^\xi = \pi_e^\xi$ and when $x^h_e = 1$, $\delta_e^\xi = 1$. Constraints (\ref{eq:stochastic_power_flow}) - (\ref{eq:stochastic_angle_limit}) ensure that the power flow in a line is 0 for out-of-service lines for every scenario. When $\delta_e = 0$, the power flow through the open line becomes 0, $P^{\xi}_e = 0$, which makes  $\theta_{i} - \theta_{j}$ unconstrained. For normal conditions, the flow is constrained by a large number $\textbf{M}$. The first-stage capacity upgrade decision $x^u_e$ is incorporated in (\ref{eq:stochastic_flow_limit}), such that if $x^u_e = 1$ then the existing line capacity will be upgraded by a factor of $\phi^u_e$, which is a known parameter. The capacity upgrade decision will still depend on whether the line is in-service ($\delta_e = 1$) or out-of-service ($\delta_e = 0$). Eq. (\ref{eq:stochastic_flow_limit}) is non-linear and hence is linearized using the Big-M method as suggested in~\cite{coelho2013linearization}.

\begin{figure}[t]
  \centering
  \includegraphics[width=0.8\linewidth]{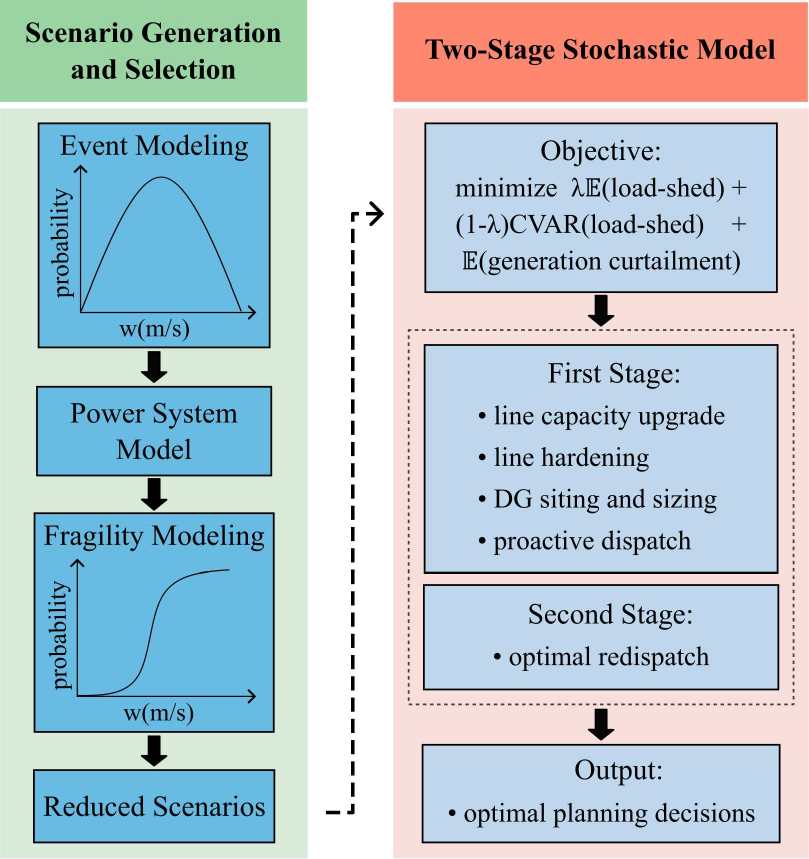}
  \caption{Overall framework of risk-based planning model.}
  \label{fig:framework}
  \vspace{-0.6cm}
\end{figure}

Constraints (\ref{eq:stochastic_generation_limit}) - (\ref{eq:gen_curtailment_limit}) limit the active power generation from a generator to its maximum capacity, load shedding limit to the maximum demand per bus, and generation curtailment to the maximum generation dispatch per generator in each scenario. The CVaR-based constraint is represented by (\ref{eq:CVaR_constraint}), which ensures that $\nu^\xi$ is greater than or equal to the load-shed at $\eta$ for each scenario $\xi \in \mathcal{E}$.


\section{Probabilistic Scenario Generation \\and Selection}\label{sec:scenario}
This section describes the scenario generation and selection procedure for resilience-based planning. It describes how an event and its impact on the power grid is modeled. A representative scenario selection method is also discussed. The overall planning framework is shown in Fig.~\ref{fig:framework}.

\subsection{Scenario Generation}

\begin{figure}[t]
  \centering
  \includegraphics[width=0.8\linewidth]{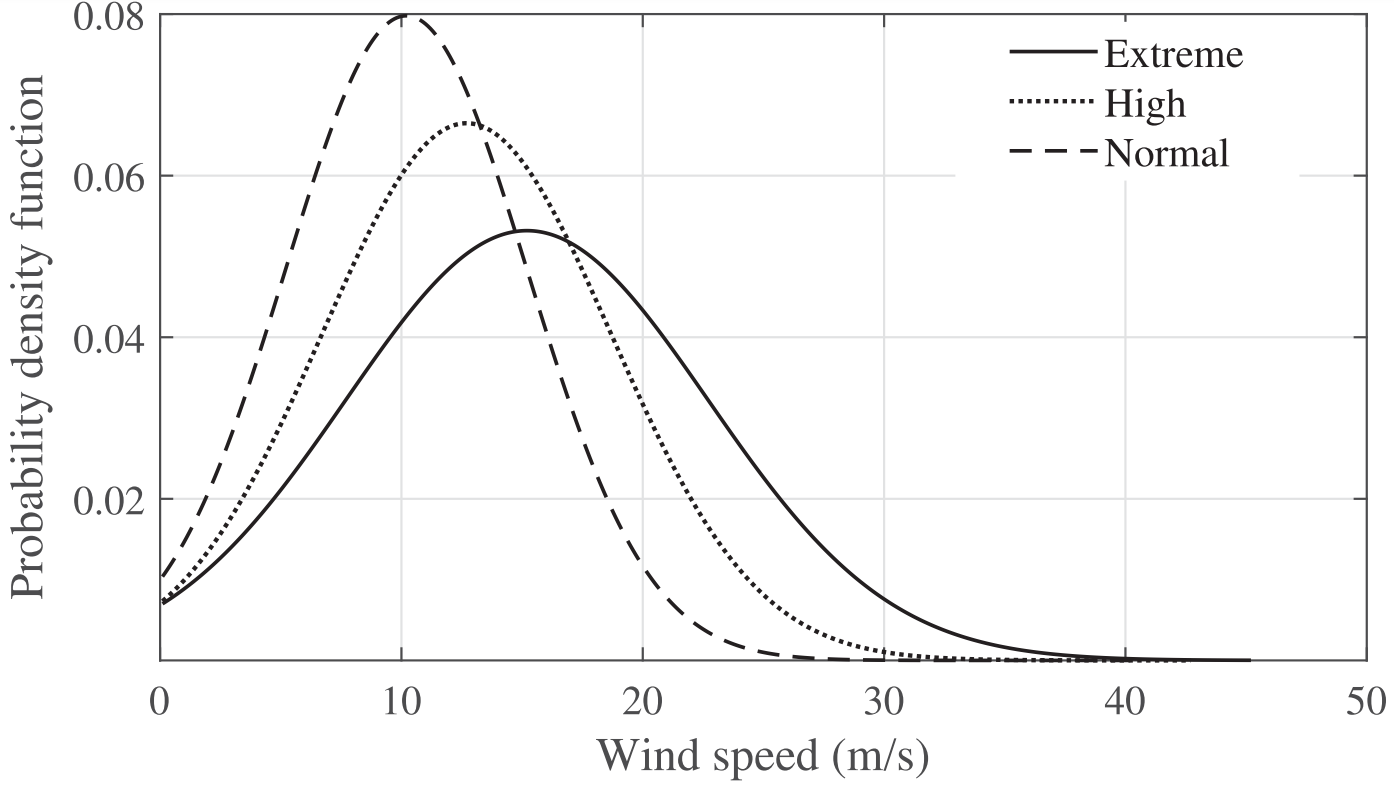}
  \caption{Regional wind profile~\cite{panteli2016power}.}
  \label{fig:regional_wind}
  \vspace{-0.4cm}
\end{figure}

An event can be defined by its probability of occurrence and the corresponding intensity. One such example of an event related to wind profile in three different regions observing extreme, high, and normal wind profiles is shown in Fig.~\ref{fig:regional_wind}. 
Component level fragility curve or prototype curve fits model can be used to assess the damage in the system due to hazards~\cite{panteli2016power}. For each wind speed, the component level fragility curve determines the operational state of the component. The fragility curve maps the wind speed to the damage probability of the component and thus can be used to determine the impacts of the event on the power system. 
Here, we use an empirical component-level fragility curve and assume that only transmission lines will be damaged due to extreme wind events. The empirical model can be replaced with known fragility models if the data is available. The mathematical relation to map wind speed to the fragility curve is given below:
$$
p_{f}(v) = \begin{cases}
    P_{\text{f}}^{n}, & \text{if } v < v_{\text{critical}} \\
    P_f(v), & \text{if } v_{\text{critical}} < v < v_{\text{collapse}} \\
    1, & \text{if } v > v_{\text{collapse}}
\end{cases}$$

where, $P_{f}(v)$ is the failure rate of the component as a function of wind speed $v$, $P_{\text{f}}^{n}$ is the failure rate at normal weather conditions, $v_{critical}$ is the wind speed at which failure probability rapidly increases and beyond $v_{collapse}$, the component fails. 

To evaluate the impact of wind-induced damages on the power system, Monte-Carlo simulations (MCS) are conducted for each wind speed to determine the system loss until the moving average loss converges. In steady-state conditions, before the system is subjected to an event, we run DCOPF on the model to mimic the normal and economic operating conditions of the generators. 
When an extreme event is realized, it is assumed that the available capacity to each generator is fixed to this pre-dispatched value as it is difficult to adjust the dispatch depending on the ramp rate and type of generator.
During or after the grid disruption due to an extreme event, the network may become disconnected, resulting in the formation of many islands. 
Loads on a particular island can get power only from generators available within the island and any mismatch between load and generation within an island is considered a load shed. In this work, we do not consider the dynamics of the grid and assume that each island is operationally stable.
The converged load loss of each wind speed is then mapped to the regional wind profile probability density function.

\subsection{Scenario Selection}
In MCS, a large number of scenarios need to be generated so that it can represent all possible circumstances. However, is is computationally challenging to incorporate all of the scenarios in the stochastic model. 
Hence, to solve the problem efficiently, scenario reduction methods are proposed to select representative scenarios from a large set of scenarios~\cite{romisch2009scenario}. 
Some other works propose sampling techniques like importance sampling~\cite{importancesampling}, stratified sampling~\cite{stratified}, and probabilistic-distance reduction~\cite{probabilisticdistance}. These methods have their own trade-off regarding solution time and quality. In this work, we use a method that is closely related to stratified sampling and distance reduction methods.

Let us consider that $N_v^\xi$ is the total number of scenarios generated for a corresponding wind speed, $v$, with  $N_v$ discrete samples of wind speed. Let $L^{avg}_v$ represent the converged MCS loss for each $v$. Then, the total number of scenarios can be represented as $\Xi=N_v^\xi\times N_v$. 
The representative scenario, $\xi_v$, is chosen such that the distance, here loss, between $\xi_v$ and average value $L^{avg}_v$ will be smallest. 
Thus, one representative scenario out of $N_v^\xi$ scenarios is chosen corresponding to each $v$. In total, we have $N_v$ representative scenarios out of total $\Xi$ scenarios. The method is described in detail in~\cite{poudyal2022risk}. Once the representative scenarios are selected, the two-stage stochastic planning model discussed in Section~\ref{sec:modeling} is then optimized for the selected range of scenarios to enhance the overall resilience of the system.
\vspace{-0.3cm}
\section{Results and Analysis}\label{sec:result}
\subsection{Test Case and Parameters}
The effectiveness of the proposed resilience-driven planning model is tested on the IEEE Reliability Test System-GMLC~\cite{barrows2019ieee}. The test system is a 73-bus system having 158 generators with a total generation capacity of 14.55 GW, 120 branches with realistic branch parameters, and a total demand of 8.55 GW. The two-stage stochastic MILP model is formulated using Pyomo~\cite{hart2017pyomo} and solved using Gurobi~\cite{gurobi2021gurobi}. Scenario generation and scenario reduction methods are implemented in MATLAB2023a, and DCOPF is solved using MATPOWER~\cite{zimmerman2010matpower}. All experiments are simulated on a PC with 16 GB RAM and Intel i7-6700 CPU @3.4GHz. The code for the overall optimization model is publicly available for reproducibility\footnote{\url{https://github.com/abodh/proactive\_resilience\_planning}}.     

The factor of generation curtailment in (\ref{eq:stochastic_objective}), $\gamma$, is selected as 0.001 to prioritize minimizing the weighted sum of expected load shed and CVaR. The confidence level for CVaR is taken as $\alpha=0.95$. The risk-neutral solution is represented by $\lambda = 0$, and the risk-averse solution is represented by $\lambda = 0.95$. The test case has different types of generator units with their corresponding ramp rates. We assume that the generators can ramp up or down to a maximum of 5 minutes, corresponding to their ramp rate. For capacity upgrade, $\phi^u_e$ is chosen as 1, which means that if a line is decided to be upgraded, then the line capacity is doubled. For each bus, the size of the DG unit is limited to 50\% of the total load, and a utilization factor of 50\% is assumed. This means that for each bus $i \in \mathcal{B}$, $P^{DG,max}_i \leq 0.25\times P^D_i$. The individual cost of other investment measures are given in Table~\ref{tab:planning_cost}. The cost of hardening a line (undergrounding) is assumed to be five times greater than the upgrade cost for the same line rating. The overall planning budget $\mathcal{C}_T^{max}$ is varied from \$0.5 billion to \$3 billion.

\begin{table}[t]
    \centering
    \caption{Cost for each planning measure.}
    \begin{tabular}{|c|c|}
    \hline
        \textbf{Planning Measure} & \textbf{Cost $\times$ (\$1 mil) } \\
    \hhline{==}
        upgrade 138 kV line     & 1.5 per mile~\cite{vea_transmission_cost}  \\
    \hline
        upgrade 230 kV line     & 1.8 per mile~\cite{pge_transmission_cost}  \\
    \hline
        harden 138 kV line & 7.5 per mile  \\
    \hline
        harden 230 kV line & 9 per mile  \\
    \hline
        DG unit installation    & 1.8 per MW~\cite{anderson2020integrating} \\
    \hline
    \end{tabular}
    \label{tab:planning_cost}
    \vspace{-0.3cm}
\end{table}

\subsection{Scenario Generation and Selection}
\begin{figure}[ht]
    \centering
    \includegraphics[width=0.8\linewidth]{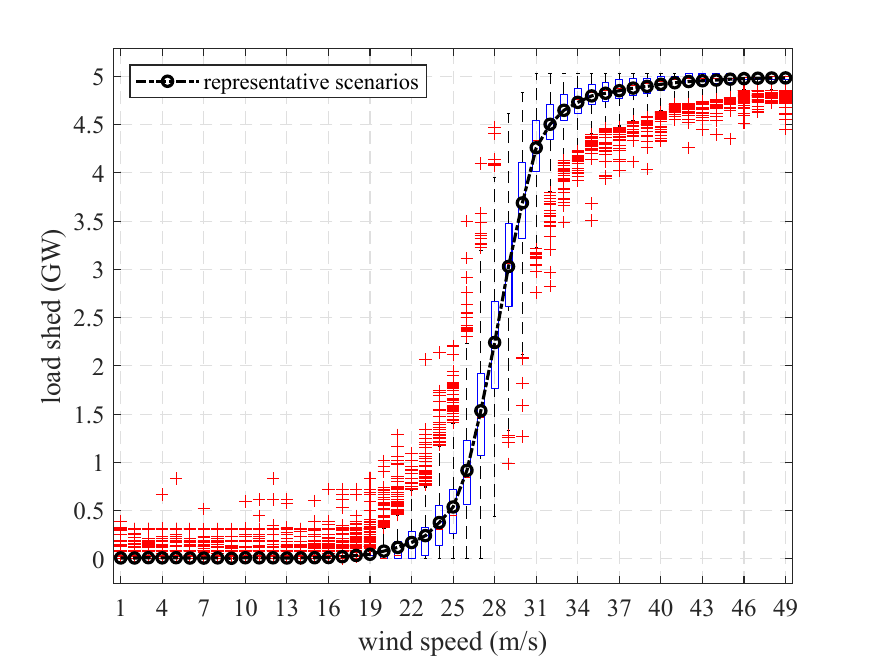}
    \caption{Loss distribution of overall scenarios for each wind speed.}
    \label{fig:all_scenarios}
    \vspace{-0.5cm}
\end{figure}

Wind events scenarios are generated using method detailed in Section III-A. We sample $N_v=49$ distinct wind speeds to represent extreme wind events, and conduct 1000 MCS trials for each wind speed considering the fragility model of transmission lines, thus resulting in a total of $\Xi=49000$ scenarios. The rationale behind choosing 1000 trials is that it ensures enough convergence for the moving average load loss in this experiment. Fig.~\ref{fig:all_scenarios} shows the box plot of load loss for all scenarios with outliers denoted by red marks. At a windspeed of $v=49~ m/s$, it is observed that the system experiences a maximum load loss of 5.033 GW, equivalent to 58.8\% of the total load. The reason for not having total blackouts even at tail-end events stems from the network topology of the transmission system and multiple generator availability at many buses. These generators create isolated islands that continue to supply power to local loads. A total of $N_v$ representative scenarios are selected based on the method described in Section III-B. The corresponding plot of reduced scenarios is shown in the black curve in Fig.~\ref{fig:all_scenarios}. It is worth mentioning that this reduction method incorporates the sampling of HILP events, which enhances its suitability for addressing resilience planning problems.

\begin{figure*}[t!]
    \centering
    \subfigure[$\hat{P}^{D}_i$ = 3384.6 MW, \% load shed = 39.58\%]{
        \includegraphics[trim={1.3cm 1cm 1.3cm 0cm},clip, width=0.48\linewidth]{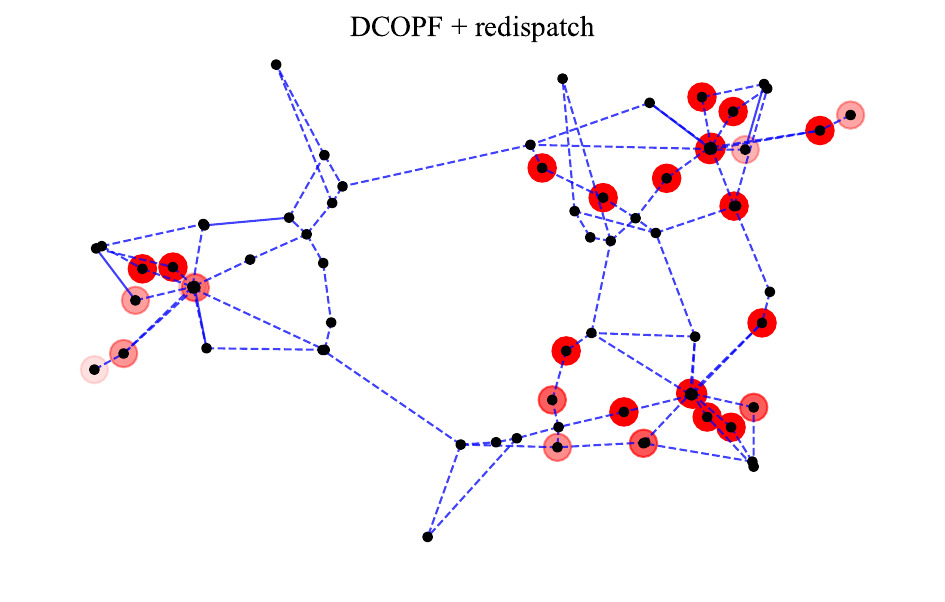}
        \label{fig:dcopf_graph}
    }\vspace{-0.1cm}
    \subfigure[$\hat{P}^{D}_i$ = 3344.6 MW, \% load shed = 39.18\%]{
        \centering
        \includegraphics[trim={1.3cm 1cm 1.3cm, 0cm},clip,width=0.48\linewidth]{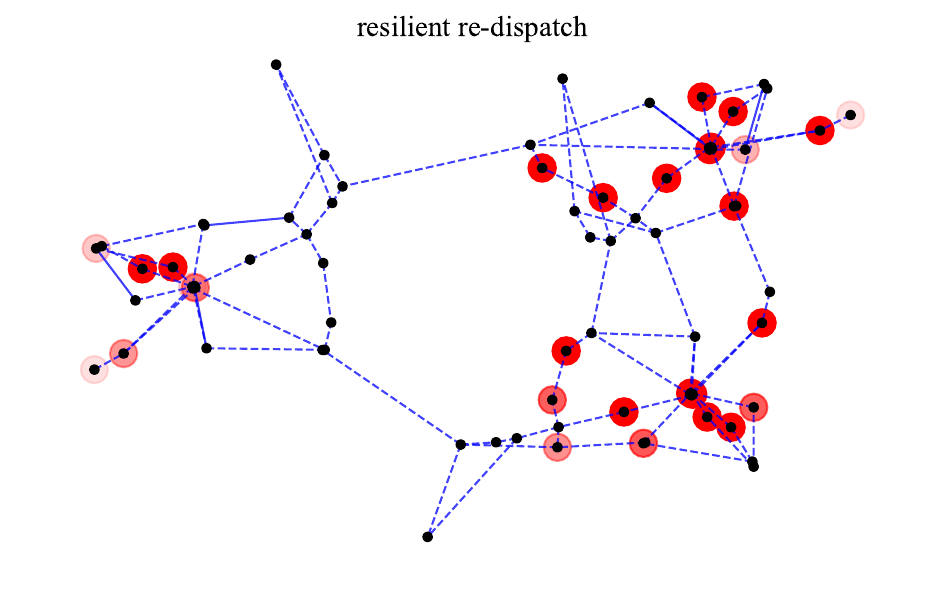}
        \label{fig:redispatch_graph}
    }\vspace{-0.1cm}
    \subfigure[$\hat{P}^{D}_i$ = 2970.9 MW, \% load shed = 34.74\%]{
        \includegraphics[trim={1.3cm 1cm 1.3cm 0cm},clip, width=0.48\linewidth]{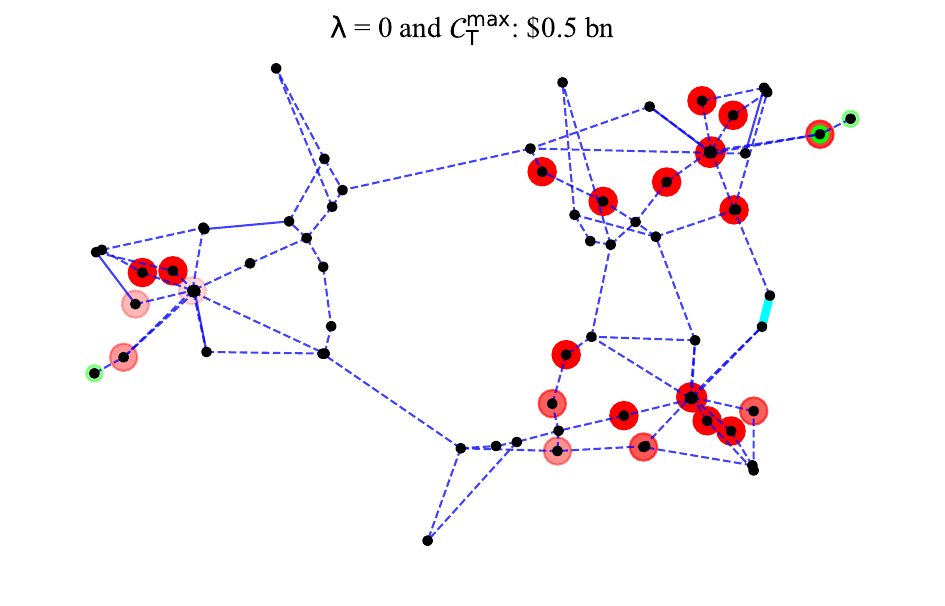}
        \label{fig:low_budget_neutral}
    }\vspace{-0.1cm}
    \subfigure[$\hat{P}^{D}_i$ = 2843.1 MW, \% load shed = 33.25\%]{
        \centering
        \includegraphics[trim={1.3cm 1cm 1.3cm, 0cm},clip,width=0.48\linewidth]{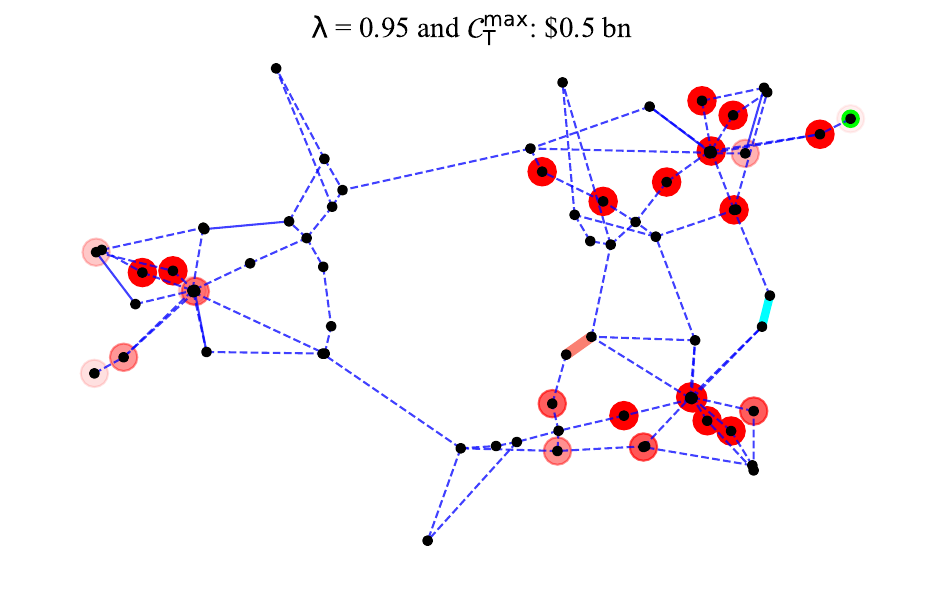}
        \label{fig:low_budget_averse}
    }\vspace{-0.1cm}
    \subfigure[$\hat{P}^{D}_i$ = 1961.55 MW, \% load shed = 22.94\%]{
        \includegraphics[trim={1.3cm 1cm 1.3cm 0cm},clip,width=0.48\linewidth]{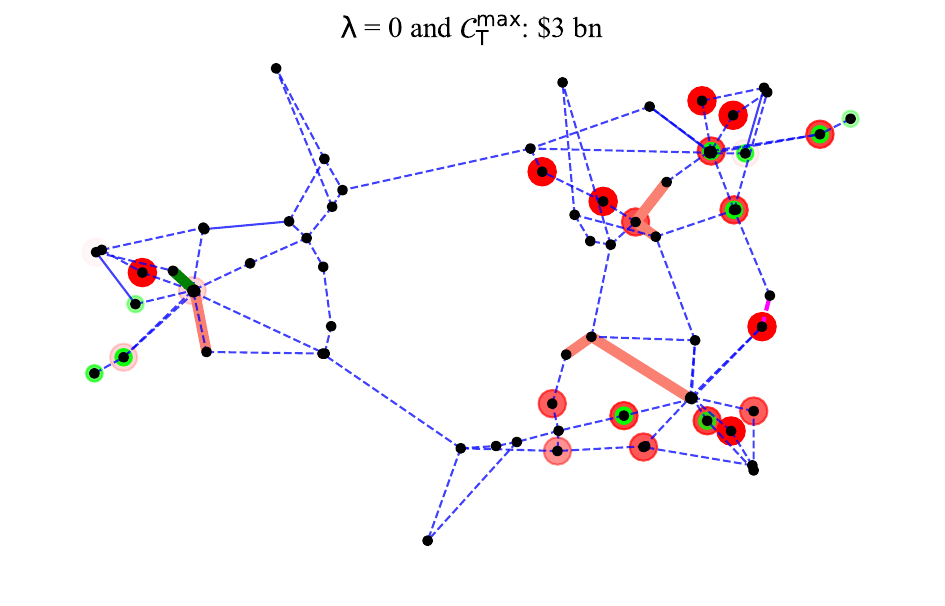}
        \label{fig:high_budget_neutral}
    }\vspace{-0.1cm}
    \subfigure[$\hat{P}^{D}_i$ = 1550.2 MW, \% load shed = 18.13\%]{
        \centering
        \includegraphics[trim={1.3cm 1cm 1.3cm 0cm},clip,width=0.48\linewidth]{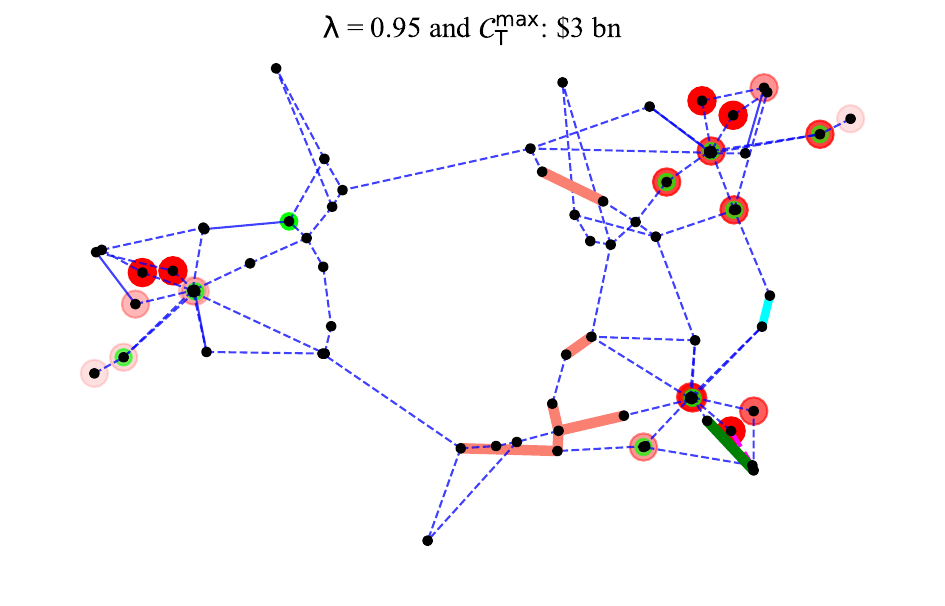}
        \label{fig:high_budget_averse}
    }\vspace{-0.1cm}
    \subfigure{
        \centering
        \includegraphics[width=\linewidth]{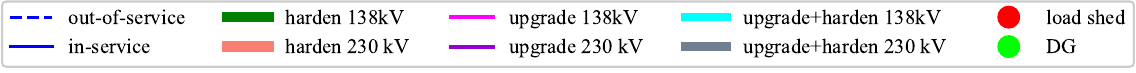}
        \label{fig:legend}
    }\vspace{-0.1cm}
    \caption{Comparison of total load shed and \% load shed for various planning strategies in a HILP scenario, $v = 41~ m/s$ and $p^\xi = 0.000672$. The planning strategies are shown in the title of each figure. The increasing opacity of red circles represents higher load shed and the increasing opacity of green circles represents buses with larger size DGs.}
    \label{fig:HILP_scenario}
\end{figure*}

\subsection{Resilience Planning}
Table.~\ref{tab:load_shed_CVAR} compares the expected load shed and CVaR for various planning strategies, budget, and risk aversion. For the base case, scenarios obtained from the scenario selection method are considered to calculate the expected load shed and CVaR. In this case, no operational decisions are carried out. For the DCOPF + re-dispatch case, generator dispatch decisions are obtained by solving the DCOPF problem without damage scenarios. Then, the dispatch decisions are fixed as the first stage dispatch decisions in the optimization model with no other planning decisions. The optimization model is then solved to consider re-dispatch decisions corresponding to each scenario. For the resilient re-dispatch case, the first-stage dispatch decisions are also obtained from the optimization model, contrary to the prior cases, where the first-stage generator dispatch decisions were fixed. The remaining cases consider resilient re-dispatch of generators along with hardening, upgrade, and DG decisions. In these cases, the maximum number of investments for each decision is fixed to 10. It can be observed that CVaR is minimum when considering all investment decisions for $\lambda=0.95, \mathcal{C}_T^{max}$ = \$ 3 bil, which is an improvement of 68.32\% from the baseline case. However, the expected load loss is minimal when considering all investment decisions for $\lambda=0, \mathcal{C}_T^{max}$ = \$ 3 bil, which is an improvement of 81.79\% from the baseline case. There is a slight trade-off observed in the expected load shed when the risk aversion increases, irrespective of budget. This is normal as the optimization model focuses on investing more to minimize the impact of HILP events, which can lower the performance slightly during expected conditions.   

\begin{table}[h]
    \centering
    \caption{Expected load shed and CVaR for various planning strategies, risk aversion and budget.}
    \begin{adjustbox}{width=\linewidth,center=\linewidth}
    \begin{tabular}{|c|c|c|c|c|c|c|}
    \hline
        \multirow{2}{*}{$\mathbf{\lambda}$} & \multirow{2}{*}{$\mathbf{\mathcal{C}_T^{max}}$} & \multirow{2}{*}{\textbf{Planning Strategy}} & \multicolumn{2}{c|}{$\mathbf{\mathbb{E}(\mathcal{F}(\textbf{x}, \mathcal{E})}$ } & \multicolumn{2}{c|}{$\mathbf{CVaR_{0.95}(\mathcal{F}(\textbf{x}, \mathcal{E})}$} \\
        \cline{4-7}
        & & & Value (MW) & Change (\%) & Value (MW) & Change (\%) \\
        
    \hhline{=======}
      -  & - & DCOPF (base case)  & 318.68  & - & 4588.07 & -  \\
    \hline
      -  & - & DCOPF + re-dispatch & 162.14  & 49.12 & 3028.22 & 34  \\
    \hline
      -  & - & resilient re-dispatch & 158.87  & 52.34 & 2966.17 & 35.35  \\
    \hline
      0  & \$0.5 bil & $N_h, N_u, N_{DG} = 10$ & 119.78  & 62.41 & 2634.85 & 42.57  \\
    \hline
      0.95  & \$0.5 bil & $N_h, N_u, N_{DG} = 10$ & 128.23  & 59.76 & 2544.01  & 44.55  \\
    \hline
      0  & \$3 bil & $N_h, N_u, N_{DG} = 10$ & \textbf{50.01}  & \textbf{81.79} & 1612.12 & 64.86  \\
    \hline
      0.95  & \$3 bil & $N_h, N_u, N_{DG} = 10$ & 77.68  & 75.62  & \textbf{1453.48} & \textbf{68.32}  \\
    \hline
    \end{tabular}
    \end{adjustbox}
    \label{tab:load_shed_CVAR}
    \vspace{-0.2cm}
\end{table}

\begin{figure*}[!t!]
    \centering
    \subfigure[risk-neutral]{
        \includegraphics[trim={1.1cm 0.1cm 1.65cm 0.4cm},clip, width=0.48\linewidth]{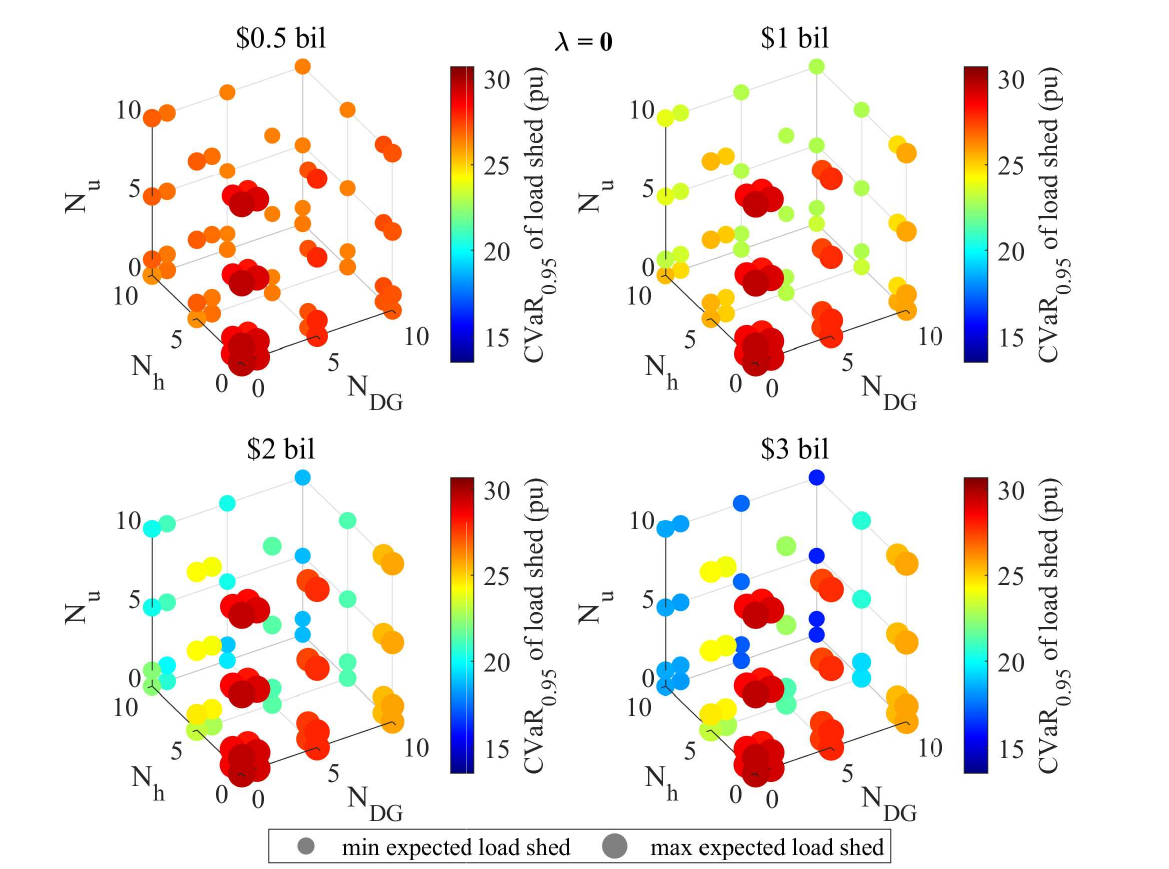}
        \label{fig:risk_neutral}
    }\vspace{-0.1cm}
    \subfigure[risk-averse]{
        \centering
        \includegraphics[trim={1.1cm 0.1cm 1.65cm 0.4cm},clip, width=0.48\linewidth]{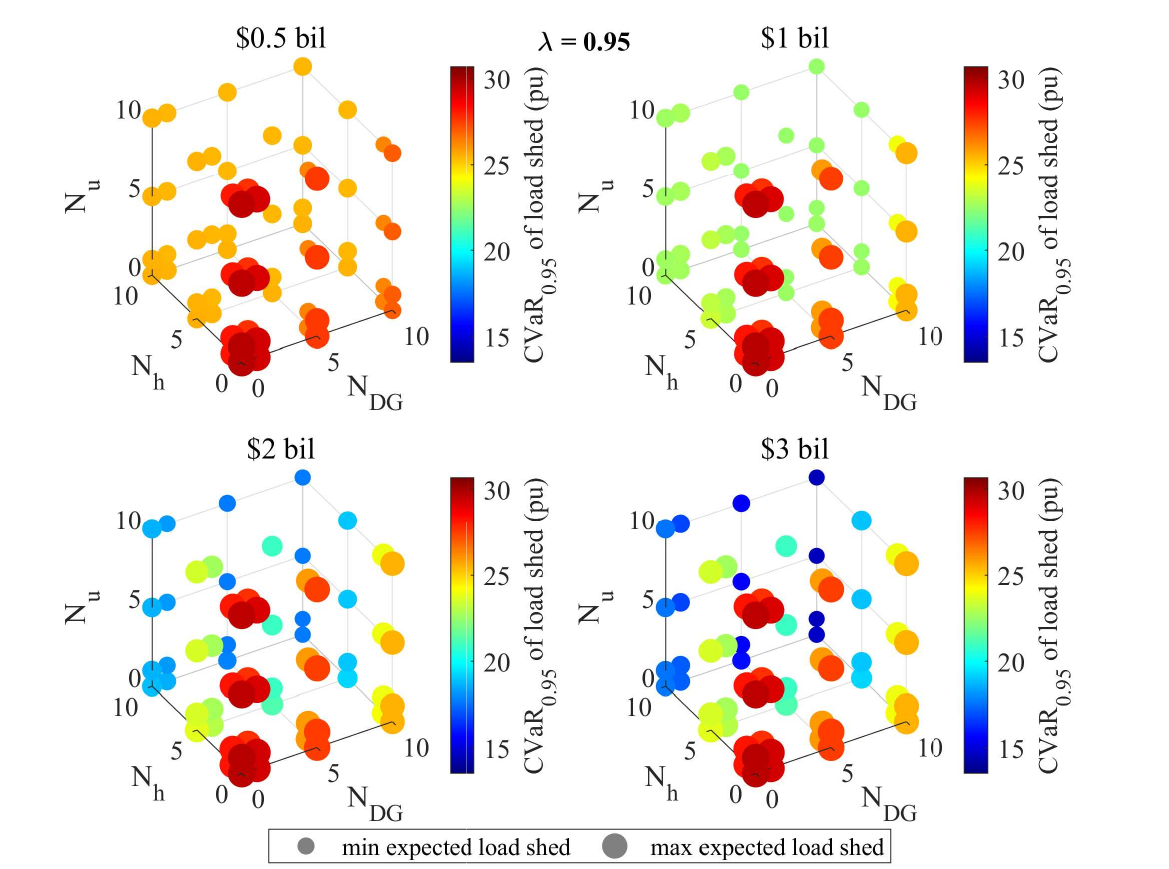}
        \label{fig:risk_averse}
    }\vspace{-0.1cm}
    \caption{Variation of expected load shed and CVaR of load shed for varying planning portfolios for risk-neutral and risk-averse decisions. The circle size represents the expected load shed for each combination, whereas the color of the circles represents CVaR. The per unit (pu) value of CVaR on the color bar reflects a system base of 100 MW.}
    \label{fig:risk_comparisons}
    \vspace{-0.5cm}
\end{figure*}

The justification above is further clarified by  Fig.~\ref{fig:HILP_scenario}, which shows the optimization decisions for a HILP scenario where $v = 41~ m/s$ and $p^\xi = 0.000672$. Although multiple lines are out-of-service in this particular scenario, the system contains several generators that can serve the load in the respective buses. However, the generators are constrained by their ramp rates and dispatch set points, which is the first-stage decision. The DCOPF + re-dispatch case is assumed as the base case here. The resilient re-dispatch model has minimal improvement over the former model in terms of load-shed minimization for this scenario. It is to be noted, however, that these decisions were made considering the entire range of scenarios, and considering all scenarios, the resilient re-dispatch model has an improvement of over 3\% in expected load shed minimization as compared to DCOPF + redispatch model, as shown in Table~\ref{tab:load_shed_CVAR}. However, it is evident that the resilient re-dispatch model does not perform well in HILP conditions. For the rest of the cases, the overall planning decisions change drastically depending upon the budget and objective. For this HILP scenario, the minimal load shed condition is achieved by the risk-averse model with \$3 billion investment. Hence, this is an important conclusion that it is essential to minimize CVaR when we want to minimize load shed for HILP scenarios. Interestingly, risk-averse models tend to invest more in hardening 230 kV lines since these lines have higher power-carrying capability and can serve more loads in case of failure of other lines.

Fig.~\ref{fig:risk_comparisons} presents a variation of expected and CVaR of load shed for various planning portfolios. 
The maximum cap on each investment decision varies within [0, 1, 5, 10], where 0 represents the case when no other planning decisions, other than generator dispatch decisions, are considered i.e., resilient re-dispatch case. For a constant budget, increasing $N_u$ has minimal impact in minimizing both the CVaR and expected load shed, whereas increasing $N_h$ has the maximum impact in all conditions. The improvement when increasing $N_{DG}$ is less than the line hardening but better than the capacity upgrade. When considering capacity upgrades only, the improvement is minimal, even if the budget is increased. In fact, increasing $N_u$ in combination with other investments shows a slight increment in risk for risk-neutral cases. However, the proper combination of line upgrade, hardening, and DG can effectively minimize the expected load shed and CVaR. Consistent with the prior results, CVaR is minimum when $\lambda = 0.95, \mathcal{C}_T^{max}$ = \$ 3 bil whereas the expected load shed is minimum when $\lambda = 0.95, \mathcal{C}_T^{max}$ = \$ 3 bil. 

\begin{figure}[ht]
    \centering    \includegraphics[width=\linewidth]{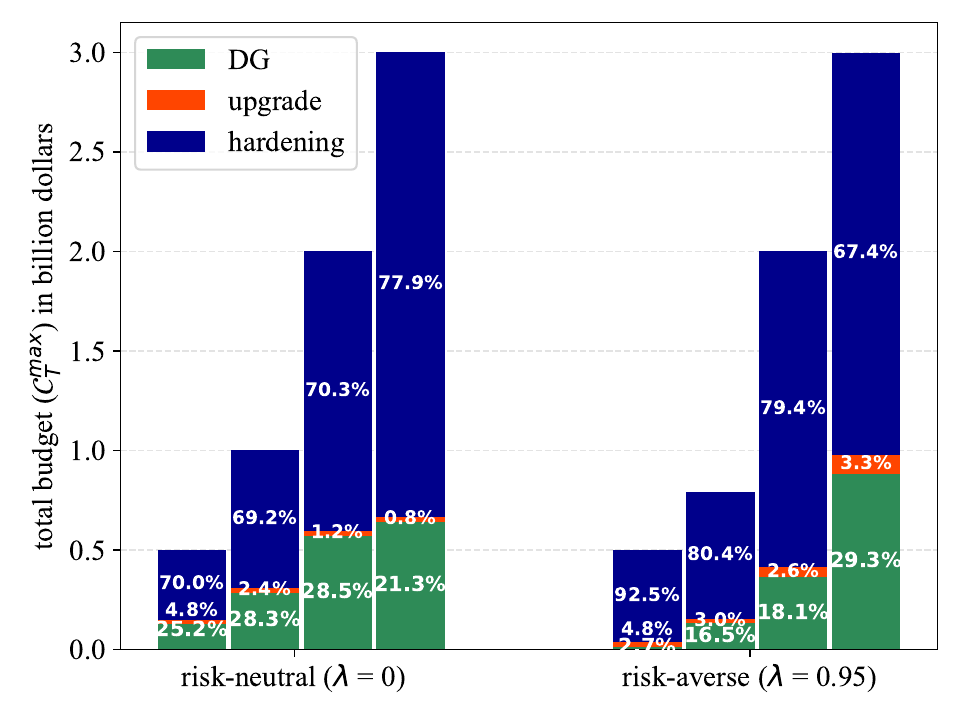}
    \caption{Budget distribution for multiple planning decisions for $N_u, N_e, N_{DG} = 10$}
    \label{fig:budget_comparison}
    \vspace{-0.5cm}
\end{figure}

The planning decisions corresponding to the $N_u, N_e, N_{DG} = 10$ are taken, and the overall budget distribution for each case is shown in Fig.~\ref{fig:budget_comparison}. Interestingly, for lower budget cases until $\mathcal{C}^{max}_T = $ \$1 billion, the risk-averse model puts a significant amount of budget in line hardening compared to other decisions. For $\mathcal{C}^{max}_T = $ \$3 billion, a considerable amount of the budget is shifted towards DGs. For the risk-neutral case, the budget proportion is almost the same for the first three budgets. However, when $\mathcal{C}^{max}_T = $ \$3 billion, the risk-neutral model increases the investment in line hardening. For almost all cases, the capacity upgrade is least adopted and is only considered once the budget for other investments has been fulfilled and the remaining budget can no longer make those investments for a given objective.

\section{Conclusion}\label{conclusion}
This work presented a resilience-driven proactive generation dispatch and multi-resource investment planning model using a two-stage stochastic optimization framework. The proposed model can identify the optimal generation dispatch, lines to harden, upgrade, and DG siting and sizing decisions, which can assist in minimizing the expected load shed and CVaR of the load shed when extreme weather scenarios are realized. The resilient re-dispatch model allows the generator to achieve a certain set point from which they can modulate within the ramp rate constraints to effectively minimize the load shedding for a range of scenarios. With detailed simulations and analysis, it was verified that the proposed method behaves considerably well as compared to DCOPF-based methods, which do not have resilience considerations. Furthermore, risk-averse decisions for the same investment budget can significantly reduce load shedding in HILP scenarios compared to other cases. It was also observed that line hardening is the most effective, and line capacity upgrade is the least effective method to minimize the load shed when considered as two independent investments. However, it was also observed that a properly designed problem with multiple resources can be effective in enhancing the resilience of the power grid against extreme weather events. A future avenue of this research is to introduce larger test case systems for scalable resilience-driven planning and analysis.

\bibliographystyle{IEEEtran}
\bibliography{ref} 

\begin{thebibliography}{10}
\providecommand{\url}[1]{#1}
\csname url@samestyle\endcsname
\providecommand{\newblock}{\relax}
\providecommand{\bibinfo}[2]{#2}
\providecommand{\BIBentrySTDinterwordspacing}{\spaceskip=0pt\relax}
\providecommand{\BIBentryALTinterwordstretchfactor}{4}
\providecommand{\BIBentryALTinterwordspacing}{\spaceskip=\fontdimen2\font plus
\BIBentryALTinterwordstretchfactor\fontdimen3\font minus \fontdimen4\font\relax}
\providecommand{\BIBforeignlanguage}[2]{{%
\expandafter\ifx\csname l@#1\endcsname\relax
\typeout{** WARNING: IEEEtran.bst: No hyphenation pattern has been}%
\typeout{** loaded for the language `#1'. Using the pattern for}%
\typeout{** the default language instead.}%
\else
\language=\csname l@#1\endcsname
\fi
#2}}
\providecommand{\BIBdecl}{\relax}
\BIBdecl

\bibitem{NOAA2023}
\BIBentryALTinterwordspacing
A.~B. SMITH, ``2022 \uppercase{U.S.} billion-dollar weather and climate disasters in historical context,'' \emph{NOAA National Centers for Environmental Information (NCEI)}, 2023. [Online]. Available: \url{https://www.climate.gov/news-features/blogs/2022-us-billion-dollar-weather-and-climate-disasters-historical-context}
\BIBentrySTDinterwordspacing

\bibitem{climatecentral2022}
\BIBentryALTinterwordspacing
{Climate Central}, ``Surging power outages and climate change,'' \emph{Climate Central}, Sep. 2022. [Online]. Available: \url{http://bit.ly/Power_Outages}
\BIBentrySTDinterwordspacing

\bibitem{ranjbarresiliency}
H.~Ranjbar, S.~H. Hosseini, and H.~Zareipour, ``Resiliency-oriented planning of transmission systems and distributed energy resources,'' \emph{IEEE Transactions on Power Systems}, vol.~36, no.~5, pp. 4114--4125, 2021.

\bibitem{datadriven}
J.~Yan, B.~Hu, K.~Xie, J.~Tang, and H.-M. Tai, ``Data-driven transmission defense planning against extreme weather events,'' \emph{IEEE Transactions on Smart Grid}, vol.~11, no.~3, pp. 2257--2270, 2019.

\bibitem{substation_line_coOptimized}
D.~Alvarado, R.~Moreno, A.~Street, M.~Panteli, P.~Mancarella, and G.~Strbac, ``Co-optimizing substation hardening and transmission expansion against earthquakes: A decision-dependent probability approach,'' \emph{IEEE Transactions on Power Systems}, vol.~38, no.~3, pp. 2058--2070, 2023.

\bibitem{reactorUsed}
A.~Soroudi, P.~Maghouli, and A.~Keane, ``{Resiliency oriented integration of DSRs in transmission networks},'' \emph{IET Generation, Transmission \& Distribution}, vol.~11, no.~8, pp. 2013--2022, 2017.

\bibitem{bynum2021proactive}
M.~Bynum, A.~Staid, B.~Arguello, A.~Castillo, B.~Knueven, C.~D. Laird, and J.-P. Watson, ``Proactive operations and investment planning via stochastic optimization to enhance power systems’ extreme weather resilience,'' \emph{Journal of Infrastructure Systems}, vol.~27, no.~2, 2021.

\bibitem{poudyal2022risk}
A.~Poudyal, S.~Poudel, and A.~Dubey, ``Risk-based active distribution system planning for resilience against extreme weather events,'' \emph{IEEE Transactions on Sustainable Energy}, vol.~14, no.~2, pp. 1178--1192, 2022.

\bibitem{barrows2019ieee}
C.~Barrows, A.~Bloom, A.~Ehlen, J.~Ik{\"a}heimo, J.~Jorgenson, D.~Krishnamurthy, J.~Lau, B.~McBennett, M.~O’Connell, E.~Preston \emph{et~al.}, ``The ieee reliability test system: A proposed 2019 update,'' \emph{IEEE Transactions on Power Systems}, vol.~35, no.~1, pp. 119--127, 2019.

\bibitem{rockafellar2000optimization}
R.~T. Rockafellar, S.~Uryasev \emph{et~al.}, ``Optimization of conditional value-at-risk,'' \emph{Journal of risk}, vol.~2, pp. 21--42, 2000.

\bibitem{coelho2013linearization}
L.~C. Coelho, ``Linearization of the product of two variables,'' \emph{Canada Research Chair in Integrated Logistics}, 2013.

\bibitem{panteli2016power}
M.~Panteli, C.~Pickering, S.~Wilkinson, R.~Dawson, and P.~Mancarella, ``Power system resilience to extreme weather: Fragility modeling, probabilistic impact assessment, and adaptation measures,'' \emph{IEEE Transactions on Power Systems}, vol.~32, no.~5, pp. 3747--3757, 2016.

\bibitem{romisch2009scenario}
W.~R{\"o}misch, ``Scenario reduction techniques in stochastic programming,'' in \emph{International Symposium on Stochastic Algorithms}.\hskip 1em plus 0.5em minus 0.4em\relax Springer, 2009, pp. 1--14.

\bibitem{importancesampling}
J.~Ekblom and J.~Blomvall, ``Importance sampling in stochastic optimization: An application to intertemporal portfolio choice,'' \emph{European Journal of Operational Research}, vol. 285, no.~1, pp. 106--119, 2020.

\bibitem{stratified}
V.~L. Parsons, ``Stratified sampling,'' \emph{Wiley StatsRef: Statistics Reference Online}, pp. 1--11, 2014.

\bibitem{probabilisticdistance}
H.~Heitsch and W.~R{\"o}misch, ``A note on scenario reduction for two-stage stochastic programs,'' \emph{Operations Research Letters}, vol.~35, no.~6, pp. 731--738, 2007.

\bibitem{hart2017pyomo}
W.~E. Hart, C.~D. Laird, J.-P. Watson, D.~L. Woodruff, G.~A. Hackebeil, B.~L. Nicholson, J.~D. Siirola \emph{et~al.}, \emph{Pyomo-optimization modeling in python}.\hskip 1em plus 0.5em minus 0.4em\relax Springer, 2017, vol.~67.

\bibitem{gurobi2021gurobi}
L.~Gurobi~Optimization, ``Gurobi optimizer reference manual,'' 2021.

\bibitem{zimmerman2010matpower}
R.~D. Zimmerman, C.~E. Murillo-S{\'a}nchez, and R.~J. Thomas, ``Matpower: Steady-state operations, planning, and analysis tools for power systems research and education,'' \emph{IEEE Transactions on power systems}, vol.~26, no.~1, pp. 12--19, 2010.

\bibitem{vea_transmission_cost}
\BIBentryALTinterwordspacing
{Valley Electric Association (VEA)}, ``{VEA 2022 Final Per Unit Cost Guide},'' \emph{{California ISO}}, 2022. [Online]. Available: \url{http://www.caiso.com/Documents/VEA2022FinalPerUnitCostGuide.xlsx}
\BIBentrySTDinterwordspacing

\bibitem{pge_transmission_cost}
\BIBentryALTinterwordspacing
{Pacific Gas and Electric (PG\&E)}, ``{PG\&E 2022 Final Per Unit Cost Guide},'' \emph{{California ISO}}, 2022. [Online]. Available: \url{http://www.caiso.com/Documents/PGE2022FinalPerUnitCostGuide.xlsx}
\BIBentrySTDinterwordspacing

\bibitem{anderson2020integrating}
K.~Anderson, X.~Li, S.~Dalvi, S.~Ericson, C.~Barrows, C.~Murphy, and E.~Hotchkiss, ``Integrating the value of electricity resilience in energy planning and operations decisions,'' \emph{IEEE Systems Journal}, vol.~15, no.~1, pp. 204--214, 2020.

\end{thebibliography}
\end{document}